# Non-invasive acquisition of fetal ECG from the maternal xyphoid process: a feasibility study in pregnant sheep and a call for open data sets

C Shen[1], MG Frasch[2,5,6], HT Wu[1,7,8], CL Herry[3], M Cao[6], A Desrochers[4], G Fecteau[4], P Burns[4]

1 Mathematics, Duke University (United States of America)
2 Obstetrics and Gynecology, University of Washington (United States of America)
3 Ottawa Hospital Research Institute (Canada)
4 Clinical Sciences, Faculty of Veterinary Medicine, Université de Montreal (Canada)
5 Centre de Recherche en Reproduction Animale (CRRA), Université de Montreal (Canada)
6 Obstetrics and Gynecology and Neurosciences, Université de Montreal (Canada)
7 Statistical Science, Duke University (United States of America)
8 Mathematics Division, National Center for Theoretical Sciences (Taiwan)

**Abstract**
*Objective:* The utility of fetal heart rate (FHR) monitoring can only be achieved with an acquisition sampling rate that preserves the underlying physiological information on the millisecond time scale (1000 Hz rather than 4 Hz). For such acquisition, fetal ECG (fECG) is required, rather than the ultrasound to derive FHR. We tested one recently developed algorithm, SAVER, and two widely applied algorithms to extract fECG from a single channel maternal ECG signal recorded over the xyphoid process rather than the routine abdominal signal. *Approach:* At 126dG, ECG was attached to near-term ewe and fetal shoulders, manubrium and xyphoid processes (n=12). FECG served as the ground-truth to which the fetal ECG signal extracted from the simultaneously-acquired maternal ECG was compared. All fetuses were in good health during surgery (pH 7.29±0.03, pO2 33.2±8.4, pCO2 56.0±7.8, O2Sat 78.3±7.6, lactate 2.8±0.6, BE -0.3±2.4). *Main result:* In all animals, single lead fECG extraction algorithm could not extract fECG from the maternal ECG signal over the xyphoid process with the F1 less than 50%. *Significance:* The applied fECG extraction algorithms might be unsuitable for the maternal ECG signal over the xyphoid process, or the latter does not contain strong enough fECG signal, although the lead is near the mother's abdomen. Fetal sheep model is widely used to mimic various fetal conditions, yet ECG recordings in a public data set form are not available to test the predictive ability of fECG and FHR. We are making this data set openly available to other researchers to foster non-invasive fECG acquisition in this animal model.

**Introduction**

The INFANT study in ~47,000 pregnancies reported no evidence of benefit on neonatal outcomes associated with the use of decision-support software in conjunction with cardiotocography (CTG) compared with CTG alone.(Group, 2017) Recent studies have shown that the true predictive ability of CTG can only be determined once it is collected at a sampling rate that preserves the underlying physiological information (i.e., 1000 Hz rather than 4 Hz).(Durosier *et al.*, 2014; Li *et al.*, 2015) This requires fetal electrocardiogram, rather than ultrasound, to derive the fetal heart rate (FHR). (Frasch *et al.*, 2017)

To address this challenge, we developed an algorithm for low-cost, portable high quality maternal and fetal ECG monitoring that could be used in human and large animal studies to record fetal ECG non-invasively and in an open format.(Wu *et al.*, 2017) (Li & Wu, 2017) While the approach has been validated in human data sets, such data sets remain scarce.

Fetal sheep model is widely used to mimic various fetal conditions, yet ECG recordings in a public data set form are not available to test the predictive ability of ECG and FHR. Part of the challenge is the requirement to perform a sterile fetal surgical instrumentation with precordial ECG leads. (Burns *et al.*, 2015) The ability to record fetal sheep ECG from the mother's surface, *i.e.*, non-invasively, would reduce the complexity of the model hopefully contributing to its wider utilization and increased availability of testable fetal ECG data sets recorded under well-controlled animal experimental conditions with translational relevance.

To explore the possibility of using fetal sheep model data, here we tested the ability of our algorithm to extract fetal ECG from a single channel maternal ECG signal recorded over the xyphoid process (maternal xyphoid ECG signal), but not the abdominal maternal ECG signal, the routine way maternal ECG is recorded in this animal model. In this study, the maternal ECG lead was placed just below the xiphoid process. Due to its proximity to the abdomen, we hypothesized that the algorithm could extract the fetal ECG from the maternal xyphoid ECG.

**Methods**

Animal care followed the guidelines of the Canadian Council on Animal Care and the approval by the University of Montreal Council on Animal Care (protocol #10-Rech-1560).

*Anesthesia and surgical procedure*
We instrumented thirteen pregnant time-dated ewes at 126 days of gestation (dGA, ~0.86 gestation) with arterial, venous and amniotic catheters and ECG electrodes.(Burns *et al.*, 2015)  Ovine singleton fetuses of mixed breed were surgically instrumented with sterile technique under general anesthesia (both ewe and fetus). In case of twin pregnancy the larger fetus was chosen based on palpating and estimating the intertemporal diameter. The total duration of the procedure was carried out in about 2 hours. Antibiotics were administered to the mother intravenously (Trimethoprim sulfadoxine 5 mg/kg) as well as to the fetus intravenously and into the amniotic cavity (ampicillin 250 mg). Amniotic fluid lost during surgery was replaced with warm saline. The catheters exteriorized through the maternal flank were secured to the back of the ewe in a plastic pouch. For the duration of the experiment the ewe was returned to the metabolic cage, where she could stand, lie and eat ad libitum while we monitored the non-anesthetized fetus without sedating the mother. During postoperative recovery antibiotic administration was continued for 3 days. Arterial blood was sampled for evaluation of maternal and fetal condition and catheters were flushed with heparinized saline to maintain patency. During surgery (once the first fetal arterial catheter was in place and before returning the fetus to the uterus) a 3 mL fetal arterial blood sample was taken for blood gas, lactate (ABL800Flex, Radiometer) and cytokine measurements. Fetal ECG was placed at first step of fetal instrumentation and recorded for the duration of surgery continuously (CED, Cambridge, U.K.). ECG was attached on both shoulders, manubrium and just below the xiphoid process (Two act as the positive and negative poles of the derivation, and the third is the ground lead) in both ewe and the fetus.

*Data acquisition*
Both maternal and fetal ECG signals were acquired simultaneously throughout the surgical instrumentation with fetal ECG serving as the ground-truth to which the fetal ECG signal extracted from the maternal ECG was compared. Maternal and fetal ECG signals were monitored continuously and digitized at 1000 Hz sampling rate with the 16 bit resolution (1902 and micro1401, both by CED, Cambridge, U.K.). (Durosier *et al.*, 2015) The database is available at doi:10.7910/DVN/NGZOPC.

*Data analysis*
The applied fetal ECG extraction algorithm for the single channel maternal xyphoid ECG signal is composed of three steps. (Li & Wu, 2017) First, the de-shape short-time Fourier transform (STFT) (Lin *et al.*, 2017) is applied to estimate the maternal instantaneous heart rate, and hence the maternal R peaks via the beat tracking algorithm. (Lin *et al.*, 2017) Second, the nonlocal Euclidean median (NLEM) is applied to recover the maternal ECG. (Li & Wu, 2017) By a direct subtraction, we obtain the *rough fetal ECG*. Finally, the fetal R peaks and hence the fetal ECG are obtained by applying the de-shape STFT, beat tracking, and the NLEM on the rough fetal ECG. In (Li & Wu, 2017), the algorithm is shown to successfully extract the fetal ECG from the maternal abdominal ECG signal. We refer the reader with interest to (Li & Wu, 2017) for details. The code is publicly available in https://sites.google.com/site/hautiengwu/home/download. See Figure 1 for a typical example for the human maternal abdominal ECG signal.

We also consider other available single channel fetal ECG extraction algorithms, particular the template subtraction (TS) approaches, including $TS_C$ proposed by Cerutti *et al.* 1986 (Cerutti *et al.*, 1986) and $TS_{PCA}$ proposed by Kanjilal *et al.* (Kanjilal *et al.*, 1997). Here, we follow the nominations proposed in (Andreotti *et al.*, 2016). The code is publicly available.

**Results**

All animals were in good health during surgery with fetal pH, pO2, pCO2, O2Sat, lactate and BE in physiological range at 7.29±0.03, 33.2±8.4, 56.0±7.8, 78.3±7.6, 2.8±0.6, -0.3±2.4, respectively. See Table 2 and 3 for details.

Four ewes carried singletons, seven ewes carried twins and one animal had triplets. Six were female and seven were male. See Table 1 for details. Maternal body weight averaged 68±6 kg and fetal body weight averaged 3.31±0.40 kg for the instrumented foetuses and did not differ from that of the twins' weight at 3.04±0.85 kg (p=0.42, two-tailed t test for unequal variances).

In all thirteen animals, we select an one minute interval of the highest quality for the analysis. Our single lead algorithm failed to extract any fetal ECG from the maternal xyphoid ECG signal. The F1 is 14.53±7.28% (range: 0-25%). $TS_C$ and $TS_{PCA}$ also failed to extract the fetal ECG, with the F1 being 12.83±6.78% (range: 0-21.6%) and 11.59±6.98% (range: 0-22.54%). See Table 4 for the case-by-case result. We found no difference in the performance of the algorithm dependent on the parity status (singleton vs. non-singleton, two-sided rank sum test, p=0.82 for the proposed method, p=0.49 for $TS_C$, and p=0.85 for $TS_{PCA}$). See Figure 2 for a typical failure example of the sheep fetal ECG derived from the maternal xyphoid ECG by the proposed algorithm. Clearly, no ECG pattern could be visualized in the rough fetal ECG.

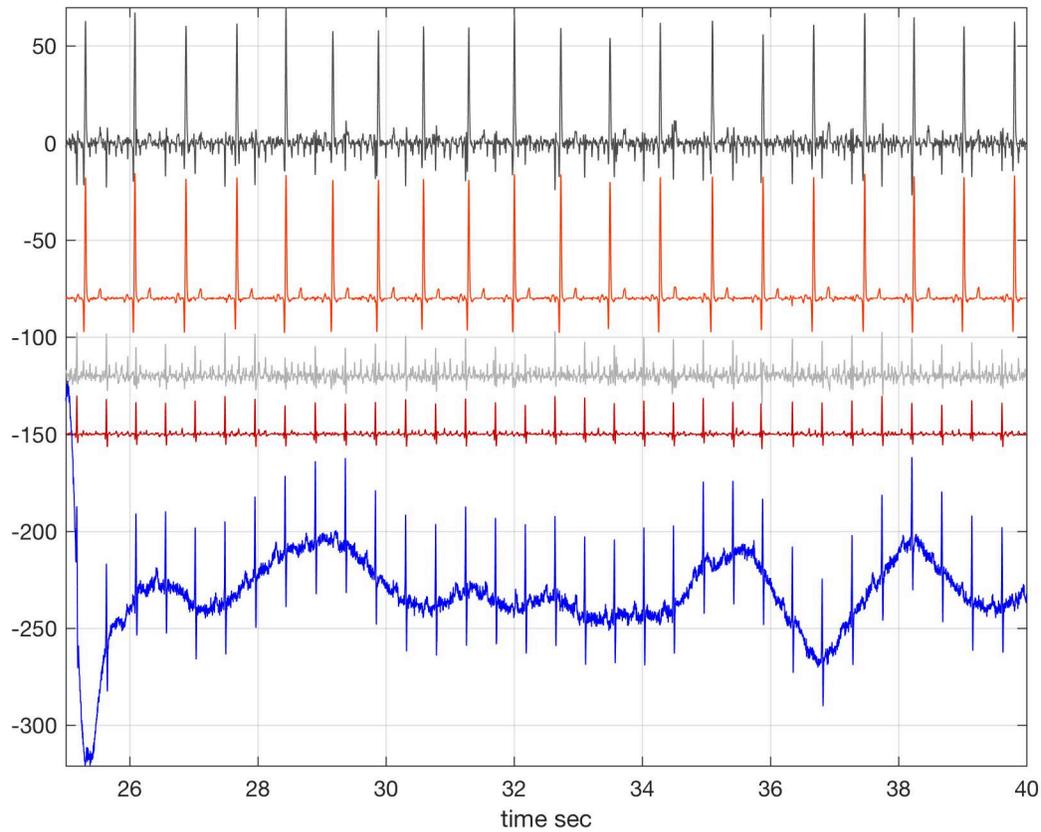

**Fig. 1.** Human aECG example, derived from the maternal abdominal ECG signal, channel1, subject 1, in the Abdominal and Direct Fetal Electrocardiogram Database (https://www.physionet.org/physiobank/database/adfecgdb/) with the proposed method. From top to bottom: the de-trended maternal abdominal ECG, the extracted maternal ECG, the rough fetal ECG, the extracted fetal ECG, and the ground truth fetal ECG. (Goldberger *et al.*, 2000), (Jezewski *et al.*, 2012)

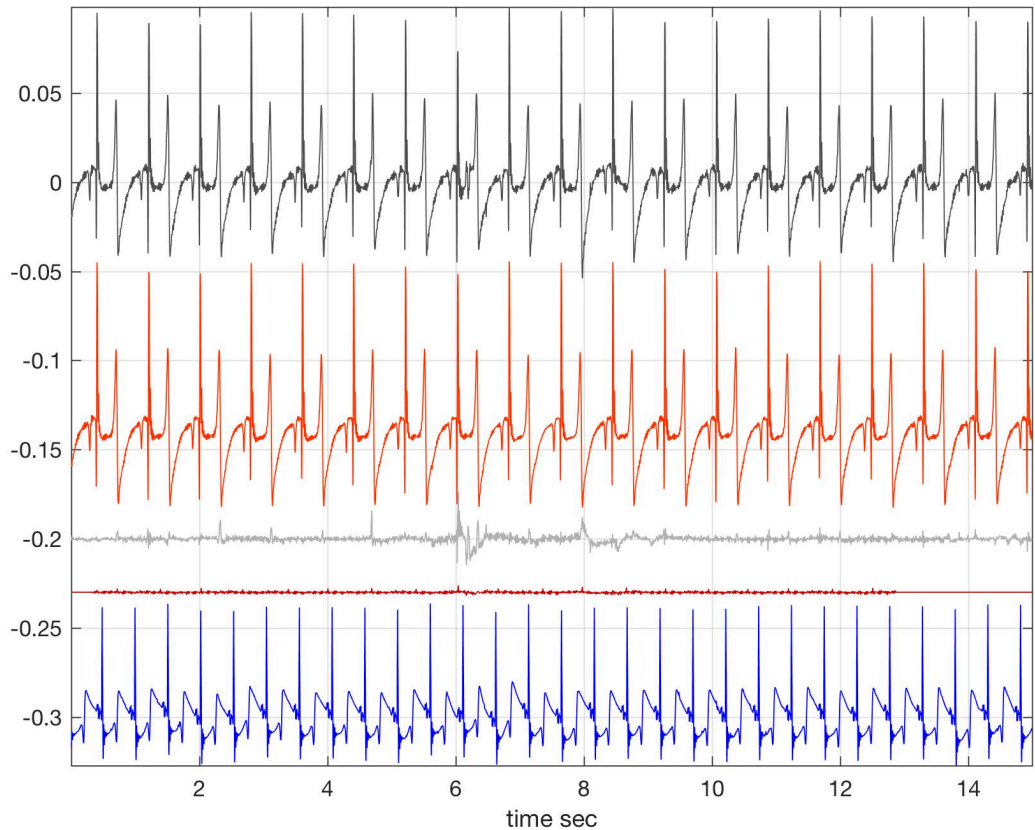

**Fig. 2.** Sheep ECG example derived from the maternal xyphoid ECG, subject 260, with the proposed method. From top to bottom: the de-trended maternal xyphoid ECG, the extracted maternal ECG, the rough fetal ECG, the extracted fetal ECG, and the ground truth fetal ECG.

**Discussion**

This paper represents an attempt of back-translation, bed-to-bench, of a long-standing tradition of deploying abdominal ECG to extract the fetal ECG signal non-invasively. (Silva *et al.*, 2013) While the standard experimental design in fetal sheep model presumes installing ECG leads on the maternal xyphoid process, it is also feasible to install the ECG leads in the abdominal area, around the incision site where the uterotomy is performed to operate on the fetus. Future studies should focus on deploying abdominal ECG leads in pregnant ewes to acquire fECG non-invasively.

Our results do not support the initial hypothesis that maternal xyphoid ECG permits the extraction of fetal ECG signal by the proposed algorithm and other widely applied single lead fetal ECG extraction algorithms $TS_C$ and $TS_{PCA}$, even when the lead is close to the uterus. There are several possible mutually non-exclusive reasons for this. First, the applied fECG extraction algorithm might not be suitable for the maternal xyphoid ECG signal. Second, the maternal xyphoid ECG signal does not contain strong enough fECG signal, although the lead is close to the mother's abdomen. Third, the sampling resolution of the data acquisition system, at 16 bits, may not suffice to capture the weaker fetal ECG signal when captured by aECG placed as far as over the maternal xyphoid. Specifically, while it is possible to sense the fetal ECG with the acquisition system at 16 bits amplitude resolution when the ECG electrode is placed on the maternal abdomen, in this study the ECG electrode was placed over the xyphoid process and the fetal ECG signal was weaker. We thus do not rule out the possibility that we can get fetal ECG from the maternal xyphoid process at higher ECG amplitude resolution. Further studies are needed to test these assumptions. Such studies could vary the following settings: testing other ECG extraction algorithms such as the scalogram approach (Khamene & Negahdaripour, 2000), the phase space approach (Karvounis *et al.*, 2007), the sequential total variation denoising approach (Lee & Lee, 2016), the adaptive noise cancellation (Zhang *et al.*, 2017), and the extended Kalman smoother (Panigrahy & Sahu, 2017); placing maternal ECG over the abdomen instead of over the xyphoid process and using data acquisition systems with amplitude resolution higher than 16 bits while recording ECG from the maternal xyphoid process. From the fetal ECG extraction perspective, note that if we have more than one ECG channel close to maternal abdomen, the blind source separation algorithms could be applied; or if we have one more ECG channel recorded from maternal thorax, we could apply adaptive filter techniques to extract the fetal ECG. (Andreotti *et al.*, 2016) However, in this study we only have single ECG signal recorded over the maternal xyphoid process, so these techniques could not be applied.

We have made this data set publicly available for those interested in fetal ECG extraction technology and invite fellow sheep model scientists to share their maternal/fetal ECG data sets in the efforts to enable a non-invasive fetal ECG acquisition in this crucial model of human physiology.

Ultimately, we hope to have highlighted the potential of launching such maternal-fetal ECG database as a benchmark for the fetal HRV analysis with high translational potential into clinical practice. The fetal sheep model is ideally suited for this endeavor, because the fetal precordial ECG quality is excellent and serves as the ground truth while permitting long-term recordings for hours to days and even weeks. As animal model, it naturally permits a more extensive control of physiological parameters than possible in a clinical setting.

**Acknowledgements**
Supported by *CIHR and FRQS (MGF) and Sloan (HTW)*

# Tables

## Table 1. Cohort characteristics

| Animal ID | Fetuses | Maternal body weight, kg | Fetal body weight, kg | Gestational age at surgery | Gender | Twin weight | Twin gender | Triplet weight | Triplet gender |
|---|---|---|---|---|---|---|---|---|---|
| 6076 | singleton | 70 | 3.69 | 127 | f | | | | |
| 6362 | singleton | 70 | 3.63 | 128 | m | | | | |
| 7329 | singleton | 57 | 3 | 127 | m | | | | |
| 7358 | singleton | 64 | 3.54 | 128 | f | | | | |
| 237 | twin | 61 | 3.46 | 127 | m | 2.32 | f | | |
| 260 | twin | 65 | 2.89 | 127 | f | 3.39 | m | | |
| 5921 | twin | 80 | 2.94 | 127 | f | 3.7 | m | | |
| 5951 | twin | 73 | 2.91 | 127 | f | 2.47 | f | | |
| 6158 | twin | 68 | 2.84 | 127 | f | 2.34 | f | | |
| 6184 | twin | 67 | 3.44 | 128 | m | 3.56 | m | | |
| 6283 | twin | 75 | 3.33 | 127 | m | 3.29 | f | | |
| 7316 | twin | 65 | 2.89 | 127 | f | 1.63 | f | | |
| 7544 | triplet | 70 | 4.1 | 128 | m | 4.1 | f | 2.9 | m |
| **Mean** | | 68 | 3.28 | 127 | | 2.98 | | | |
| **SD** | | 6 | 0.40 | 0 | | 0.81 | | | |

Table 2. Fetal arterial blood gas during surgery

| Animal ID | pH | paCO$_2$, mmHg | paO$_2$, mmHg | Hb, g/dL | Hct | O$_2$Sat, % | O$_2$Content | Glucose, mg/dL | Lactate, mmol/L | HCO$_3$ | BE | Na | K | Ca | Cl |
|---:|---:|---:|---:|---:|---:|---:|---:|---:|---:|---:|---:|---:|---:|---:|---:|
| 6076 | 7.25 | 71.5 | 35.7 | 14.3 | 43.8 | 82.4 | 15.0 | 24.0 | 2.5 | 29.4 | 3.3 | 135.0 | 3.7 | 1.6 | 99.0 |
| 6362 | 7.24 | 58.1 | 34.6 | 12.5 | 38.3 | 79.7 | 13.4 | 22.0 | 2.8 | 23.7 | -2.2 | 133.0 | 3.6 | 1.4 | 101.0 |
| 7329 | 7.33 | 41.1 | 58.9 | 10.5 | 32.3 | 97.8 | 13.8 | 26.0 | 3.8 | 20.4 | -4.2 | 131.0 | 3.7 | 1.5 | 104.0 |
| 7358 | 7.26 | 53.4 | 32.1 | 11.7 | 36.0 | 80.6 | 12.2 | 21.0 | 2.1 | 22.5 | -3.1 | 138.0 | 3.7 | 1.5 | 106.0 |
| 237 | 7.32 | 55.0 | 33.3 | 12.3 | 37.7 | 84.2 | 13.9 | 18.0 | 2.2 | 26.5 | 1.5 | 133.0 | 3.4 | 1.5 | 98.0 |
| 260 | 7.28 | 57.1 | 28.7 | 12.3 | 38.0 | 73.3 | 12.2 | 17.0 | 2.5 | 25.2 | -0.3 | 136.0 | 3.4 | 1.5 | 106.0 |
| 5921 | 7.29 | 61.8 | 28.9 | 12.8 | 39.3 | 70.3 | 12.2 | 19.0 | 3.3 | 28.0 | 2.6 | 136.0 | 3.2 | 1.8 | 101.0 |
| 5951 | 7.33 | 56.2 | 27.4 | 13.8 | 42.5 | 73.2 | 13.7 | 13.0 | 2.2 | 27.6 | 2.7 | 138.0 | 3.8 | 1.5 | 103.0 |
| 6158 | 7.29 | 57.0 | 32.3 | 11.6 | 35.6 | 78.5 | 12.3 | 18.0 | 2.3 | 25.6 | 0.2 | 137.0 | 3.2 | 1.5 | 104.0 |
| 6184 | 7.30 | 48.4 | 30.1 | 12.5 | 38.6 | 73.8 | 12.5 | 22.0 | 3.5 | 22.2 | -2.9 | 139.0 | 3.7 | 1.7 | 110.0 |
| 6283 | 7.26 | 62.8 | 29.5 | 12.3 | 37.8 | 72.4 | 12.0 | 24.0 | 2.7 | 26.2 | 0.3 | 134.0 | 3.4 | 1.5 | 103.0 |
| 7316 | 7.35 | 45.9 | 24.2 | 13.2 | 40.5 | 69.9 | 12.4 | 23.0 | 3.6 | 23.8 | -0.7 | 134.0 | 3.6 | 1.5 | 101.0 |
| 7544 | 7.26 | 59.9 | 36.2 | 13.1 | 40.1 | 81.3 | 14.3 | 26.0 | 3.3 | 25.0 | -0.8 | 137.0 | 3.8 | 1.3 | 104.0 |
| Mean | 7.29 | 56.0 | 33.2 | 12.5 | 38.5 | 78.3 | 13.1 | 21.0 | 2.8 | 25.1 | -0.3 | 135.5 | 3.6 | 1.5 | 103.1 |
| SD | 0.03 | 7.8 | 8.4 | 1.0 | 3.0 | 7.6 | 1.0 | 3.8 | 0.6 | 2.5 | 2.4 | 2.4 | 0.2 | 0.1 | 3.2 |

Table 3. Maternal arterial blood gas during surgery

| Animal ID | pH | paCO$_2$, mmHg | paO$_2$, mmHg | Hb, g/dL | Hct | O$_2$Sat, % | O$_2$Content | HCO3 | BE | Na | K | Ca | Cl |
|---|---|---|---|---|---|---|---|---|---|---|---|---|---|
| 6076 | 7.35 | 54.7 | 434.0 | 8.5 | 26.4 | 66.0 | 1.2 | 28.1 | 3.4 | 143.0 | 3.4 | 1.1 | 106.0 |
| 6362 | 7.39 | 47.8 | 500.0 | 9.3 | 28.8 | 68.0 | 1.0 | 26.9 | 2.8 | 140.0 | 3.3 | 1.0 | 105.0 |
| 7329 | 7.51 | 33.0 | 485.0 | 8.8 | 27.2 | 67.0 | 1.2 | 24.7 | 2.3 | 138.0 | 2.9 | 1.0 | 105.0 |
| 7358 | 7.40 | 41.3 | 455.0 | 8.3 | 25.8 | 57.0 | 1.2 | 24.3 | 0.5 | 142.0 | 3.6 | 1.2 | 105.0 |
| 237 | 7.34 | 46.1 | 330.0 | 9.6 | 29.7 | 37.0 | 1.1 | 23.5 | -1.1 | 139.0 | 3.6 | 1.1 | 103.0 |
| 260 | 7.41 | 45.8 | 349.0 | 8.5 | 26.4 | 52.0 | 0.6 | 27.3 | 3.5 | 142.0 | 4.1 | 1.1 | 108.0 |
| 5921 | 7.44 | 43.5 | 436.0 | 8.3 | 25.8 | 61.0 | 1.5 | 27.8 | 4.5 | 143.0 | 3.3 | 1.1 | 105.0 |
| 5951 | 7.44 | 39.3 | 373.0 | 8.1 | 25.3 | 51.0 | 1.8 | 25.0 | 1.7 | 145.0 | 3.3 | 1.0 | 109.0 |
| 6158 | 7.39 | 44.0 | 443.0 | 8.3 | 25.9 | 55.0 | 0.9 | 25.2 | 1.3 | 145.0 | 3.3 | 1.1 | 110.0 |
| 6184 | 7.44 | 35.2 | 440.0 | 9.3 | 28.7 | 54.0 | 1.7 | 22.3 | -0.9 | 147.0 | 3.3 | 1.0 | 114.0 |
| 6283 | 7.36 | 46.2 | 207.0 | 8.9 | 27.5 | 76.0 | 1.1 | 24.4 | 0.1 | 141.0 | 3.2 | 1.1 | 107.0 |
| 7316 | 7.50 | 32.7 | 355.0 | 8.3 | 25.8 | 79.0 | 2.4 | 24.1 | 1.7 | 140.0 | 2.9 | 1.1 | 104.0 |
| 7544 | 7.35 | 48.0 | 456.0 | 7.7 | 23.9 | 64.0 | 1.1 | 24.8 | 0.3 | 144.0 | 3.5 | 1.0 | 109.0 |
| Mean | 7.41 | 42.9 | 404.8 | 8.6 | 26.7 | 60.5 | 1.3 | 25.3 | 1.5 | 142.2 | 3.4 | 1.1 | 106.9 |
| SD | 0.06 | 6.4 | 79.8 | 0.5 | 1.6 | 11.3 | 0.5 | 1.8 | 1.7 | 2.6 | 0.3 | 0.0 | 3.0 |

**Table 4. The fetal ECG extraction result, where the fetal R peak detection measured in F1 is reported.**

|        | Time intervals | Proposed | TS_c  | TS_PCA |
|--------|----------------|----------|-------|--------|
| #237:  | 4000:4060      | 19.47    | 19.79 | 19.75  |
| #260:  | 4000:4060      | 6.35     | 12.29 | 14.53  |
| #5921: | 6300:6360      | 4.69     | 21.6  | 22.54  |
| #5951  | 3000:3060      | 14.46    | 0     | 0      |
| #6076  | 3500:3560      | 0        | 6.76  | 1.52   |
| #6158  | 2900:2960      | 11.27    | 15.38 | 3.85   |
| #6184  | 5000:5060      | 18.18    | 0     | 5.46   |
| #6283  | 3000:3060      | 13.9     | 15.6  | 16.51  |
| #6362  | 3000:3060      | 22.89    | 16.28 | 11.11  |
| #7316  | 1000:1060      | 17.39    | 12.87 | 13.71  |
| #7329  | 1000:1060      | 19.1     | 17.35 | 15.08  |
| #7358  | 1000:1060      | 16.24    | 16.49 | 15.46  |
| #7544  | 300:360        | 25       | 12.44 | 11.11  |
| **Mean** |              | 14.53    | 12.83 | 11.59  |
| **SD** |                | 7.28     | 6.78  | 6.98   |